\documentclass[a4paper,11pt]{article}
\usepackage{pos}

\title{Evolving the COLA software library}

\author*[a]{Waseem Kamleh}

\affiliation[a]{Centre for the Subatomic Structure of Matter, Department of Physics,\\ The University of Adelaide, SA 5005, Australia}

\emailAdd{waseem.kamleh@adelaide.edu.au}

\abstract{COLA is a software library for lattice QCD, written in a combination of modern Fortran and C/C++. Intel and NVIDIA have dominated the HPC domain in the years leading up to the exascale era, but the status quo has changed with the arrival of Frontier and other AMD-based systems in the supercomputing Top 500. Setonix is a next generation HPE Cray EX system hosted at the Pawsey Supercomputing Centre in Perth, Australia. Setonix features AMD EPYC CPUs and AMD Instinct GPUs. This report describes some of my experiences in evolving COLA to adapt to the current hardware landscape.}

\FullConference{%
  The 39th International Symposium on Lattice Field Theory (Lattice2022),\\
  8-13 August, 2022 \\
  Bonn, Germany 
}

\begin{document}
\maketitle

The first supercomputer in the world to achieve a HPL benchmark greater than 1 exaflop/s was \href{https://www.ornl.gov/news/ornl-celebrates-launch-frontier-worlds-fastest-supercomputer}{Frontier, based at the Oak Ridge National Laboratory.} Frontier is a HPE Cray EX system featuring AMD EPYC ``Trento'' cores and Radeon Instinct MI250X accelerators. Topping both the \href{https://www.top500.org/lists/top500/2022/06/}{June} and \href{https://www.top500.org/lists/top500/2022/11/}{November Top 500} lists in 2022 with an $R_{\rm max}$ score of 1.102 exaflop/s, Frontier displaced the Fugaku supercomputer at RIKEN, which features a bespoke chipset based on the ARM architecture. The previous crown holder was Summit, another Oak Ridge entry, but with NVIDIA accelerators and an IBM Power system. The diversity of hardware architectures on leading systems represents one of the challenges of high performance computing. Research groups typically have allocations on a variety of machines, and with the recent resurgence of AMD the range of target architectures has only widened. It is not just platform portability for scientific codes that is essential, but also performance portability.

The challenges of portability in the context of scientific computing are not new of course, but there are aspects of the contemporary architecture ecosystem that are distinct. Historically, there have always been a range of CPU chipsets deployed at HPC facilities. Performant scientific codes are typically written in C/C++ or Fortran. These languages are not tied to a specific vendor,  and it is reasonable to expect that the compilers for these languages are available on any system. In this sense, platform portability for CPU codes presents a relatively low barrier for code development. For performance portability, due to tight language restrictions around aliasing, Fortran compilers developed by hardware vendors such as Intel and Cray have traditionally been very successful at generating optimised code. With regard C/C++, the use of architecture specific intrinsic has often been required to get the most benefit from specific processor features (such as vectorisation).

COLA is a custom in-house code that I began developing in Fortran as a graduate student. The code is more or less in a constant state of change, either evolving to adapt to new challenges or expanding to add new features. Key algorithms are solvers for linear systems and eigenmodes~\cite{Kalkreuter:1995mm}, and gauge field generation with Hybrid Monte Carlo~\cite{Duane:1987de}. The latter features a number of variants such as the RHMC algorithm~\cite{Kennedy:1998cu} and a selection of filtering techniques~\cite{Kamleh:2011dc,Haar:2016bwe,Haar:2018jjd}. Specific physics features are tailored to the CSSM lattice research programme, for which the COLA software library has formed the computational foundation for some time~\cite{Kusterer:2001vk,Kamleh:2001ff,Kamleh:2004xk,Boinepalli:2004fz,Kamleh:2004aw,Lasscock:2005tt,Hedditch:2005zf,Boinpolli:2007zz,Hedditch:2007ex,Kamleh:2007ud,Lasscock:2007ce,Kamleh:2007bd,Mahbub:2009nr,Mahbub:2009aa,Mahbub:2010jz,Mahbub:2010me,Mahbub:2010rm,Roberts:2010cz,Menadue:2011pd,OMalley:2011aa,Roberts:2012tp,Mahbub:2012ri,Owen:2012ts,Mahbub:2013ala,Stokes:2013fgw,Roberts:2013ipa,Trewartha:2013qga,Primer:2013pva,Mahbub:2013bba,Roberts:2013oea,Stokes:2013oaa,Thomas:2014tda,Hall:2014uca,Owen:2015gva,Kiratidis:2015vpa,Trewartha:2015nna,Owen:2015fra,Trewartha:2015ida,Liu:2015ktc,Dragos:2016rtx,Liu:2016uzk,Kiratidis:2016hda,Hall:2016kou,Trewartha:2017ive,Bignell:2018acn,Biddle:2018dtc,Wu:2018tvt,Stokes:2018emx,Stokes:2019zdd,Bignell:2019vpy,Virgili:2019shg,Biddle:2019gke,Bignell:2020xkf,Bignell:2020dze,CSSMQCDSFUKQCD:2021rvs,Virgili:2022ybm,Biddle:2022zgw,Biddle:2022acd}.

The fastest machine in Australia in the November 2022 Top 500 is Setonix, ranked at \#15 with a HPL score of 27 petaflop/s and hosted at the Pawsey Supercomputing Centre in Perth. Similar to Frontier, Setonix is based on the AMD EPYC platform and also features the Radeon MI250X accelerators. Building upon the energy efficiency of these accelerators, \href{https://pawsey.org.au/australias-setonix-ranking/}{Setonix ranks at \#4 in the Green 500 list} based on performance per Watt, 2 places higher than Frontier at \#6. As the world moves towards a carbon-neutral future, there has been a developing focus on optimising the energy efficiency~\cite{DBLP:journals/corr/abs-2110-09987} of traditionally power hungry high performance computing systems.

To accompany the release of this novel architecture to the Australian supercomputing scene, the \href{https://pawsey.org.au/pacer/}{Pawsey Centre for Extreme Scale Readiness (PaCER)} scheme was created. Several projects were chosen to partner with Pawsey to optimise codes and workflows for the next generation of supercomputers. The CSSM is partnered with the PaCER scheme via one of these projects, \emph{Emergent phenomena revealed in subatomic matter}. User community initiatives with similar goals have accompanied the launch of other recent AMD-based systems.

\begin{figure}[t]
  \centering%
  \includegraphics[width=0.7\textwidth]{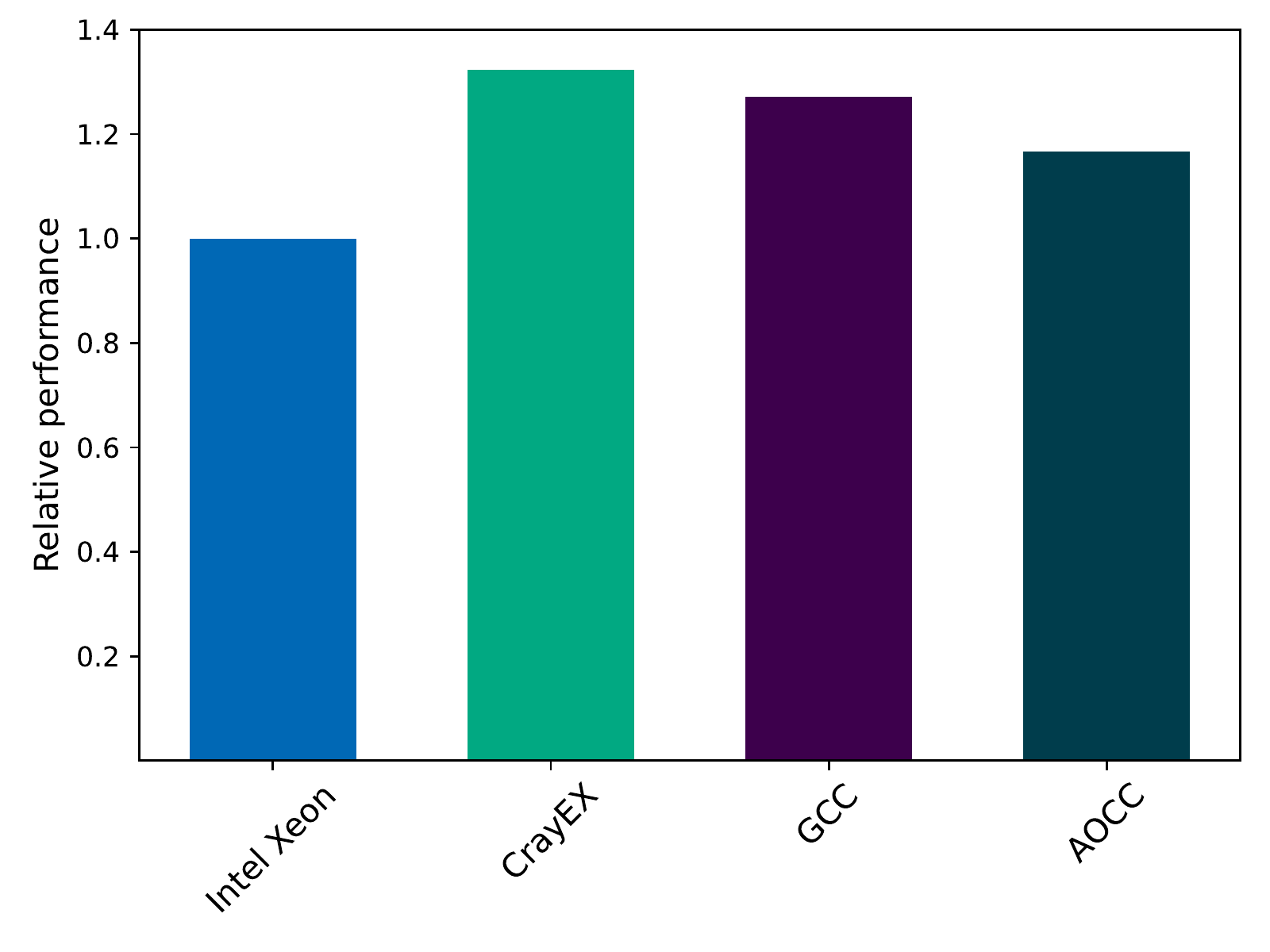}
  \caption{Performance of the COLA fermion matrix on the AMD EPYC ``Milan'' CPU with the CrayEX, GCC, and AOCC Fortran compilers, measured relative to the Xeon ``Cascade Lake'' performance with the Intel Fortran compiler.}
  \label{fig:cpuperf}
\end{figure}

As the Fortran components of COLA do not use any vendor or architecture specific features, adapting the software to the AMD EPYC platform was fairly straightforward. The enforcement of standard compliance does vary between the different compilers, requiring some minor changes for code which has primarily evolved on the Intel platform.
There are three programming environments available on Setonix, namely Cray EX, GNU GCC, and AOCC (which is based on LLVM). Figure~\ref{fig:cpuperf} shows the relative performance of the fermion matrix code under these three compilers on a single dual-socket ``Milan'' CPU node with a total of 128 cores, as compared to the Intel compiler on a dual-socket Xeon ``Skylake'' node with 48 cores. As would be expected given the effort that they put into their Fortran compiler optimisation, Cray performs the best of the three AMD programming environments, closely followed by the GNU compiler. At the time of writing, modern Fortran support within the AOCC programming environment should be considered as preliminary.

GPU acceleration for the COLA fermion matrix inverter was first introduced via NVIDIA CUDA C/C++ around the time that the Fermi architecture was released. This mixed-language approach persists to this day. As the CPU aspects of the code are written in modern Fortran, I utilise the interoperability provided by the intrinsic \emph{ISO C Binding} module to interface with the GPU-accelerated routines that are implemented in CUDA C/C++. A consequence of this approach is that for many of the key algorithms in COLA there exist two independent implementations -- one in Fortran for CPUs and one in CUDA for GPUs. The ability to cross-check the two implementations as a form of validation has proved very beneficial during code development. Significant amounts of utility code reuse are also realised, with the Fortran code that handles the reading of runtime parameters, the input and output of data, and the initialisation of the MPI process topology being common to both the CPU and GPU implementations.

Lattice codes have a relatively low flops/byte ratio and as such are typically memory bandwidth limited. The intrinsic geometric parallelism of the lattice maps naturally onto the massively parallel nature of the GPU platform. Whilst on a CPU a subvolume of the lattice would typically be mapped to a single core, on a GPU each lattice site is mapped to a single thread. The latency hiding capabilities of the GPU execution model coupled with the high computational power means that opportunites to perform additional computation in order to reduce traffic from global memory generally result in an overall speedup of the code. Similarly, the use of mixed-precision techniques also results in a significant benefit~\cite{Clark:2009wm}. Historically, the amount of device memory available was relatively small compared to the size of a fermion field. The available device memory on GPU accelerators has increased significantly with successive generations, such that limiting the number of vectors stored is less of a concern than it once was (similarly for register pressure).

\begin{figure}[t]
  \centering%
  \includegraphics[width=0.7\textwidth]{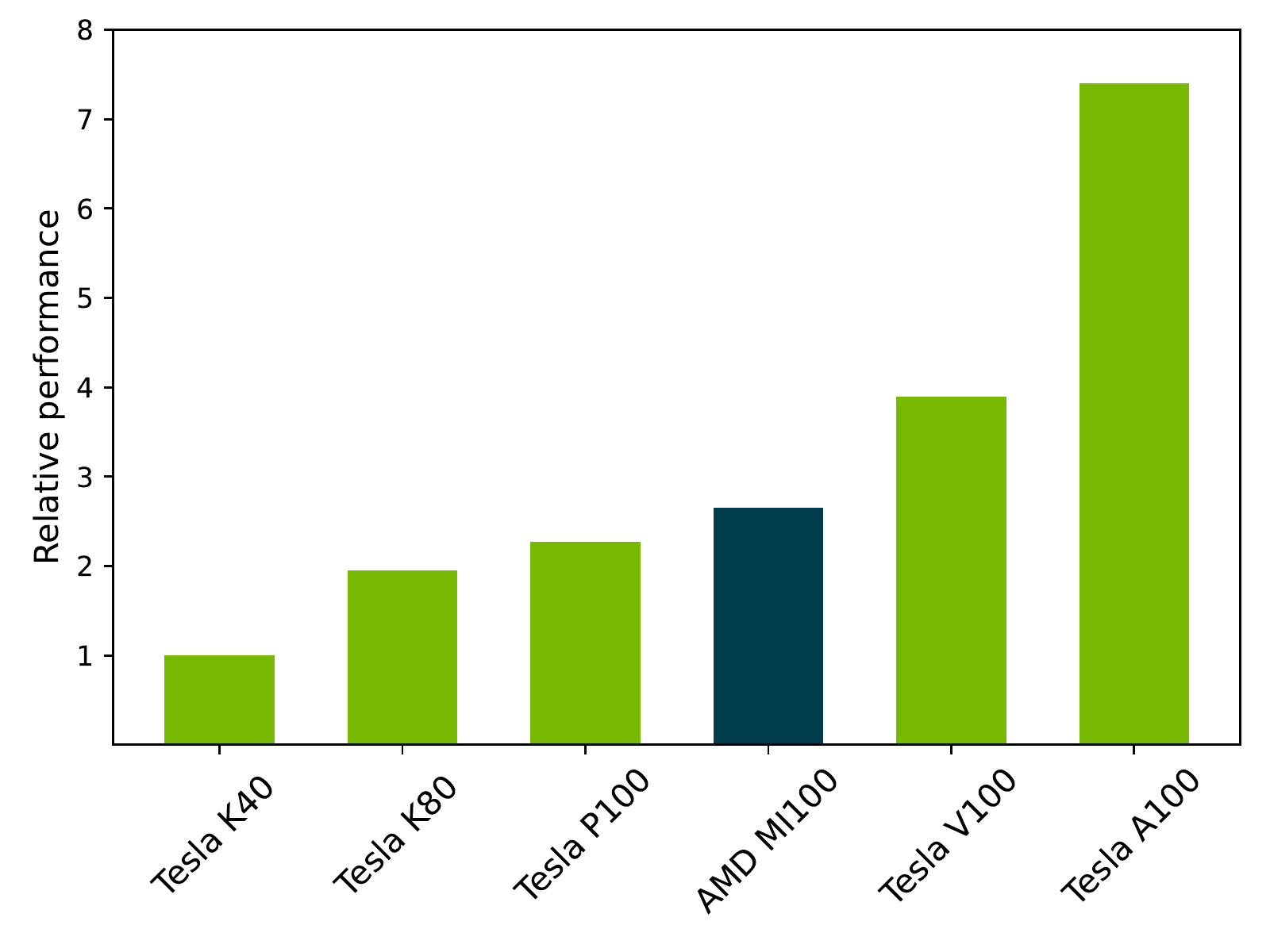}
  \caption{Performance of the COLA GPU-accelerated conjugate gradient inverter on various device architectures, measured relative to the Tesla K40 performance. Results are shown for various NVIDIA Tesla cards  with generations ranging from Kepler to Ampere, and the AMD MI100 accelerator.}
  \label{fig:gpuperf}
\end{figure}

Figure~\ref{fig:gpuperf} shows the relative performance of the GPU-accelerated COLA conjugate gradient inverter on a variety of GPU architectures, using the NVIDIA Tesla K40 as a reference. Starting with Kepler, NVIDIA results are provided for generations up to Ampere with the A100. There is a single data point for AMD Instinct, the MI100. A mixed-precision solver is used for the benchmark, with 32-bit precision for the inner iterations and 64-bit precision for the outer solver.

To run on the MI100 architecture the CUDA components of COLA were converted to AMD's Heterogeneous Interface for Portability (HIP). The HIP SDK provides scripts that automated much of the process of converting from CUDA code, although some manual effort was required to complete the conversion. The HIP compiler can target both NVIDIA and AMD accelerators, and on the Volta platform at least seemed to provide equivalent performance to \emph{nvcc}. At the time of writing, the MI250X was not yet available to the author for benchmarking. Setonix will be fully launched in early 2023, when the MI250X partition becomes available to users. While platform portability has been demonstrated for the COLA software, it will be interesting to see what can be achieved in terms of performance on the MI250X.

It seems that one of the aims of HIP is to mirror the functionality provided by CUDA, though this approach necessarily means there is a delay between a CUDA feature release and the appearance of the HIP equivalent. The dominance of CUDA for GPU-accelerated codes can be attributed to the fact that NVIDIA was the first vendor to successfully bring to market devices that targeted the HPC community. Arguably, their success would not have been possible without the large amount of effort put into developing the CUDA programming environment. Many researchers have implemented their code in CUDA due to a combination of the rapid maturity it achieved relative to other programming models (such as OpenCL), and the near-monopoly NVIDIA has had on accelerated HPC systems for many years.

AMD has had recent success at the hardware level with the launch of high profile HPC systems such as Frontier in the US, Lumi in Europe, and Setonix in Australia. They must now ensure that their programming environment rapidly achieves maturity in order to sustain momentum. It is also interesting to note that Intel is attempting to (re-)enter into the accelerator space with the Xe HPC platform.

This again raises the issue of (platform and performance) portability for accelerated computing. CPUs can generally be targeted by Fortran and C/C++ code in a platform-independent manner whilst maintaining performance (of course, platform-specific optimisations can always improve upon these). For accelerators, it is a different story. Vendor-specific programming environments are typically required to get the best performance. Developing divergent branches of the same code adds significant overhead to an activity that is already at a premium in the academic research environment, where time well-spent must necessarily translate into the \emph{de facto} currency of the field.

In an ideal world we would treat the accelerator space in much the same way as we treat the traditional compute space. That is, through the establishment of platform-independent programming models with agreed upon standards that compiler providers implement to target their respective hardware platforms. Arguably, the natural way for this to proceed would be to have vendor-agnostic accelerated programming extensions to C/C++ and Fortran. Fortran has included intrinsic parallel computing features since the 1990s, and accelerator-based programming would seem like a natural extension. Language standards tend to evolve fairly slowly however, so in the short term we must look elsewhere. There are vendor-led candidates for cross-platform heterogeneous programming such as NVIDIA's OpenACC, AMD HIP, and Intel's DPC++ for oneAPI. Of course, these can be expected to perform well on the vendor's respective hardware, but support for (and performance on) the competitors hardware is not guaranteed. There are also a number of open candidates for heterogeneous programming such as OpenCL, SYCL, and Kokkos. The extent to which the various candidates above provide performance portability is being investigated~\cite{Bach:2012iw,Boyle:2017gzg,Bonati:2018wqj,Cali:2021adj,Yamaguchi:2022feu}. Future work will explore this question in the context of continuing to evolve the COLA software library.

\section*{Acknowledgements}

The author is supported by the Pawsey Supercomputing Centre through
the Pawsey Centre for Extreme Scale Readiness (PaCER) program. This
work is supported by the Australian Research Council through
Grants No. DP190102215 and DP210103706.


\begin{thebibliography}{10}

\bibitem{Kalkreuter:1995mm}
T.~Kalkreuter and H.~Simma, \emph{{An Accelerated conjugate gradient algorithm
  to compute low lying eigenvalues: A Study for the Dirac operator in SU(2)
  lattice QCD}},
  \href{https://doi.org/10.1016/0010-4655(95)00126-3}{\emph{Comput. Phys.
  Commun.} {\bfseries 93} (1996) 33}
  [\href{https://arxiv.org/abs/hep-lat/9507023}{{\ttfamily hep-lat/9507023}}].

\bibitem{Duane:1987de}
S.~Duane, A.D.~Kennedy, B.J.~Pendleton and D.~Roweth, \emph{{Hybrid Monte
  Carlo}}, \href{https://doi.org/10.1016/0370-2693(87)91197-X}{\emph{Phys.
  Lett. B} {\bfseries 195} (1987) 216}.

\bibitem{Kennedy:1998cu}
A.D.~Kennedy, I.~Horvath and S.~Sint, \emph{{A New exact method for dynamical
  fermion computations with nonlocal actions}},
  \href{https://doi.org/10.1016/S0920-5632(99)85217-7}{\emph{Nucl. Phys. B
  Proc. Suppl.} {\bfseries 73} (1999) 834}
  [\href{https://arxiv.org/abs/hep-lat/9809092}{{\ttfamily hep-lat/9809092}}].

\bibitem{Kamleh:2011dc}
W.~Kamleh and M.~Peardon, \emph{{Polynomial Filtered HMC: An Algorithm for
  lattice QCD with dynamical quarks}},
  \href{https://doi.org/10.1016/j.cpc.2012.05.002}{\emph{Comput. Phys. Commun.}
  {\bfseries 183} (2012) 1993}
  [\href{https://arxiv.org/abs/1106.5625}{{\ttfamily 1106.5625}}].

\bibitem{Haar:2016bwe}
T.~Haar, W.~Kamleh, J.~Zanotti and Y.~Nakamura, \emph{{Applying polynomial
  filtering to mass preconditioned Hybrid Monte Carlo}},
  \href{https://doi.org/10.1016/j.cpc.2017.02.020}{\emph{Comput. Phys. Commun.}
  {\bfseries 215} (2017) 113}
  [\href{https://arxiv.org/abs/1609.02652}{{\ttfamily 1609.02652}}].

\bibitem{Haar:2018jjd}
T.~Haar, W.~Kamleh, J.~Zanotti and Y.~Nakamura, \emph{{Single flavour
  optimisations to Hybrid Monte Carlo}},
  \href{https://doi.org/10.1016/j.cpc.2018.12.009}{\emph{Comput. Phys. Commun.}
  {\bfseries 238} (2019) 111}
  [\href{https://arxiv.org/abs/1806.04350}{{\ttfamily 1806.04350}}].

\bibitem{Kusterer:2001vk}
D.-J.~Kusterer, J.~Hedditch, W.~Kamleh, D.B.~Leinweber and A.G.~Williams,
  \emph{{Low lying eigenmodes of the Wilson-Dirac operator and correlations
  with topological objects}},
  \href{https://doi.org/10.1016/S0550-3213(02)00070-6}{\emph{Nucl. Phys. B}
  {\bfseries 628} (2002) 253}
  [\href{https://arxiv.org/abs/hep-lat/0111029}{{\ttfamily hep-lat/0111029}}].

\bibitem{Kamleh:2001ff}
W.~Kamleh, D.H.~Adams, D.B.~Leinweber and A.G.~Williams, \emph{{Accelerated
  overlap fermions}},
  \href{https://doi.org/10.1103/PhysRevD.66.014501}{\emph{Phys. Rev. D}
  {\bfseries 66} (2002) 014501}
  [\href{https://arxiv.org/abs/hep-lat/0112041}{{\ttfamily hep-lat/0112041}}].

\bibitem{Kamleh:2004xk}
W.~Kamleh, D.B.~Leinweber and A.G.~Williams, \emph{{Hybrid Monte Carlo with fat
  link fermion actions}},
  \href{https://doi.org/10.1103/PhysRevD.70.014502}{\emph{Phys. Rev. D}
  {\bfseries 70} (2004) 014502}
  [\href{https://arxiv.org/abs/hep-lat/0403019}{{\ttfamily hep-lat/0403019}}].

\bibitem{Boinepalli:2004fz}
S.~Boinepalli, W.~Kamleh, D.B.~Leinweber, A.G.~Williams and J.M.~Zanotti,
  \emph{{Improved chiral properties of FLIC fermions}},
  \href{https://doi.org/10.1016/j.physletb.2005.04.050}{\emph{Phys. Lett. B}
  {\bfseries 616} (2005) 196}
  [\href{https://arxiv.org/abs/hep-lat/0405026}{{\ttfamily hep-lat/0405026}}].

\bibitem{Kamleh:2004aw}
W.~Kamleh, P.O.~Bowman, D.B.~Leinweber, A.G.~Williams and J.~Zhang, \emph{{The
  fat link irrelevant clover overlap quark propagator}},
  \href{https://doi.org/10.1103/PhysRevD.71.094507}{\emph{Phys. Rev. D}
  {\bfseries 71} (2005) 094507}
  [\href{https://arxiv.org/abs/hep-lat/0412022}{{\ttfamily hep-lat/0412022}}].

\bibitem{Lasscock:2005tt}
B.G.~Lasscock, J.N.~Hedditch, W.~Kamleh, D.B.~Leinweber, W.~Melnitchouk,
  A.W.~Thomas et~al., \emph{{Search for the pentaquark resonance signature in
  lattice QCD}}, \href{https://doi.org/10.1103/PhysRevD.72.014502}{\emph{Phys.
  Rev. D} {\bfseries 72} (2005) 014502}
  [\href{https://arxiv.org/abs/hep-lat/0503008}{{\ttfamily hep-lat/0503008}}].

\bibitem{Hedditch:2005zf}
J.N.~Hedditch, W.~Kamleh, B.G.~Lasscock, D.B.~Leinweber, A.G.~Williams and
  J.M.~Zanotti, \emph{{1-+ exotic meson at light quark masses}},
  \href{https://doi.org/10.1103/PhysRevD.72.114507}{\emph{Phys. Rev. D}
  {\bfseries 72} (2005) 114507}
  [\href{https://arxiv.org/abs/hep-lat/0509106}{{\ttfamily hep-lat/0509106}}].

\bibitem{Boinpolli:2007zz}
S.~Boinpolli, P.O.~Bowman, U.M.~Heller, W.~Kamleh, J.N.~Hedditch, B.G.~Lasscock
  et~al., \emph{{Some recent lattice QCD results from the CSSM}},
  \href{https://doi.org/10.1142/S0217751X07038402}{\emph{Int. J. Mod. Phys. A}
  {\bfseries 22} (2007) 5053}.

\bibitem{Hedditch:2007ex}
J.N.~Hedditch, W.~Kamleh, B.G.~Lasscock, D.B.~Leinweber, A.G.~Williams and
  J.M.~Zanotti, \emph{{Pseudoscalar and vector meson form-factors from lattice
  QCD}}, \href{https://doi.org/10.1103/PhysRevD.75.094504}{\emph{Phys. Rev. D}
  {\bfseries 75} (2007) 094504}
  [\href{https://arxiv.org/abs/hep-lat/0703014}{{\ttfamily hep-lat/0703014}}].

\bibitem{Kamleh:2007ud}
W.~Kamleh, P.O.~Bowman, D.B.~Leinweber, A.G.~Williams and J.~Zhang,
  \emph{{Unquenching effects in the quark and gluon propagator}},
  \href{https://doi.org/10.1103/PhysRevD.76.094501}{\emph{Phys. Rev. D}
  {\bfseries 76} (2007) 094501}
  [\href{https://arxiv.org/abs/0705.4129}{{\ttfamily 0705.4129}}].

\bibitem{Lasscock:2007ce}
B.G.~Lasscock, J.N.~Hedditch, W.~Kamleh, D.B.~Leinweber, W.~Melnitchouk,
  A.G.~Williams et~al., \emph{{Even parity excitations of the nucleon in
  lattice QCD}}, \href{https://doi.org/10.1103/PhysRevD.76.054510}{\emph{Phys.
  Rev. D} {\bfseries 76} (2007) 054510}
  [\href{https://arxiv.org/abs/0705.0861}{{\ttfamily 0705.0861}}].

\bibitem{Kamleh:2007bd}
W.~Kamleh, B.~Lasscock, D.B.~Leinweber and A.G.~Williams, \emph{{Scaling
  analysis of FLIC fermion actions}},
  \href{https://doi.org/10.1103/PhysRevD.77.014507}{\emph{Phys. Rev. D}
  {\bfseries 77} (2008) 014507}
  [\href{https://arxiv.org/abs/0709.1531}{{\ttfamily 0709.1531}}].

\bibitem{Mahbub:2009nr}
M.S.~Mahbub, A.~O.~Cais, W.~Kamleh, B.G.~Lasscock, D.B.~Leinweber and
  A.G.~Williams, \emph{{Isolating Excited States of the Nucleon in Lattice
  QCD}}, \href{https://doi.org/10.1103/PhysRevD.80.054507}{\emph{Phys. Rev. D}
  {\bfseries 80} (2009) 054507}
  [\href{https://arxiv.org/abs/0905.3616}{{\ttfamily 0905.3616}}].

\bibitem{Mahbub:2009aa}
M.S.~Mahbub, A.O.~Cais, W.~Kamleh, B.G.~Lasscock, D.B.~Leinweber and
  A.G.~Williams, \emph{{Isolating the Roper Resonance in Lattice QCD}},
  \href{https://doi.org/10.1016/j.physletb.2009.07.063}{\emph{Phys. Lett. B}
  {\bfseries 679} (2009) 418}
  [\href{https://arxiv.org/abs/0906.5433}{{\ttfamily 0906.5433}}].

\bibitem{Mahbub:2010jz}
M.S.~Mahbub, A.O.~Cais, W.~Kamleh, D.B.~Leinweber and A.G.~Williams,
  \emph{{Positive-parity Excited-states of the Nucleon in Quenched Lattice
  QCD}}, \href{https://doi.org/10.1103/PhysRevD.82.094504}{\emph{Phys. Rev. D}
  {\bfseries 82} (2010) 094504}
  [\href{https://arxiv.org/abs/1004.5455}{{\ttfamily 1004.5455}}].

\bibitem{Mahbub:2010me}
M.S.~Mahbub, W.~Kamleh, D.B.~Leinweber, A.~O~Cais and A.G.~Williams,
  \emph{{Ordering of Spin-$\frac{1}{2}$ Excitations of the Nucleon in Lattice
  QCD}}, \href{https://doi.org/10.1016/j.physletb.2010.08.049}{\emph{Phys.
  Lett. B} {\bfseries 693} (2010) 351}
  [\href{https://arxiv.org/abs/1007.4871}{{\ttfamily 1007.4871}}].

\bibitem{Mahbub:2010rm}
{\scshape CSSM Lattice} collaboration, \emph{{Roper Resonance in 2+1 Flavor
  QCD}}, \href{https://doi.org/10.1016/j.physletb.2011.12.048}{\emph{Phys.
  Lett. B} {\bfseries 707} (2012) 389}
  [\href{https://arxiv.org/abs/1011.5724}{{\ttfamily 1011.5724}}].

\bibitem{Roberts:2010cz}
D.S.~Roberts, P.O.~Bowman, W.~Kamleh and D.B.~Leinweber, \emph{{Wave Functions
  of the Proton Ground State in the Presence of a Uniform Background Magnetic
  Field in Lattice QCD}},
  \href{https://doi.org/10.1103/PhysRevD.83.094504}{\emph{Phys. Rev. D}
  {\bfseries 83} (2011) 094504}
  [\href{https://arxiv.org/abs/1011.1975}{{\ttfamily 1011.1975}}].

\bibitem{Menadue:2011pd}
B.J.~Menadue, W.~Kamleh, D.B.~Leinweber and M.S.~Mahbub, \emph{{Isolating the
  $\Lambda(1405)$ in Lattice QCD}},
  \href{https://doi.org/10.1103/PhysRevLett.108.112001}{\emph{Phys. Rev. Lett.}
  {\bfseries 108} (2012) 112001}
  [\href{https://arxiv.org/abs/1109.6716}{{\ttfamily 1109.6716}}].

\bibitem{OMalley:2011aa}
E.-A.~O'Malley, W.~Kamleh, D.~Leinweber and P.~Moran, \emph{{SU(3) centre
  vortices underpin confinement and dynamical chiral symmetry breaking}},
  \href{https://doi.org/10.1103/PhysRevD.86.054503}{\emph{Phys. Rev. D}
  {\bfseries 86} (2012) 054503}
  [\href{https://arxiv.org/abs/1112.2490}{{\ttfamily 1112.2490}}].

\bibitem{Roberts:2012tp}
D.S.~Roberts, W.~Kamleh, D.B.~Leinweber, M.S.~Mahbub and B.J.~Menadue,
  \emph{{Accessing High Momentum States In Lattice QCD}},
  \href{https://doi.org/10.1103/PhysRevD.86.074504}{\emph{Phys. Rev. D}
  {\bfseries 86} (2012) 074504}
  [\href{https://arxiv.org/abs/1206.5891}{{\ttfamily 1206.5891}}].

\bibitem{Mahbub:2012ri}
M.S.~Mahbub, W.~Kamleh, D.B.~Leinweber, P.J.~Moran and A.G.~Williams,
  \emph{{Low-lying Odd-parity States of the Nucleon in Lattice QCD}},
  \href{https://doi.org/10.1103/PhysRevD.87.011501}{\emph{Phys. Rev. D}
  {\bfseries 87} (2013) 011501}
  [\href{https://arxiv.org/abs/1209.0240}{{\ttfamily 1209.0240}}].

\bibitem{Owen:2012ts}
B.J.~Owen, J.~Dragos, W.~Kamleh, D.B.~Leinweber, M.S.~Mahbub, B.J.~Menadue
  et~al., \emph{{Variational Approach to the Calculation of gA}},
  \href{https://doi.org/10.1016/j.physletb.2013.04.063}{\emph{Phys. Lett. B}
  {\bfseries 723} (2013) 217}
  [\href{https://arxiv.org/abs/1212.4668}{{\ttfamily 1212.4668}}].

\bibitem{Mahbub:2013ala}
M.S.~Mahbub, W.~Kamleh, D.B.~Leinweber, P.J.~Moran and A.G.~Williams,
  \emph{{Structure and Flow of the Nucleon Eigenstates in Lattice QCD}},
  \href{https://doi.org/10.1103/PhysRevD.87.094506}{\emph{Phys. Rev. D}
  {\bfseries 87} (2013) 094506}
  [\href{https://arxiv.org/abs/1302.2987}{{\ttfamily 1302.2987}}].

\bibitem{Stokes:2013fgw}
F.M.~Stokes, W.~Kamleh, D.B.~Leinweber, M.S.~Mahbub, B.J.~Menadue and
  B.J.~Owen, \emph{{Parity-expanded variational analysis for nonzero
  momentum}}, \href{https://doi.org/10.1103/PhysRevD.92.114506}{\emph{Phys.
  Rev. D} {\bfseries 92} (2015) 114506}
  [\href{https://arxiv.org/abs/1302.4152}{{\ttfamily 1302.4152}}].

\bibitem{Roberts:2013ipa}
D.S.~Roberts, W.~Kamleh and D.B.~Leinweber, \emph{{Wave Function of the Roper
  from Lattice QCD}},
  \href{https://doi.org/10.1016/j.physletb.2013.06.056}{\emph{Phys. Lett. B}
  {\bfseries 725} (2013) 164}
  [\href{https://arxiv.org/abs/1304.0325}{{\ttfamily 1304.0325}}].

\bibitem{Trewartha:2013qga}
A.~Trewartha, W.~Kamleh, D.~Leinweber and D.S.~Roberts, \emph{{Quark
  Propagation in the Instantons of Lattice QCD}},
  \href{https://doi.org/10.1103/PhysRevD.88.034501}{\emph{Phys. Rev. D}
  {\bfseries 88} (2013) 034501}
  [\href{https://arxiv.org/abs/1306.3283}{{\ttfamily 1306.3283}}].

\bibitem{Primer:2013pva}
T.~Primer, W.~Kamleh, D.~Leinweber and M.~Burkardt, \emph{{Magnetic properties
  of the nucleon in a uniform background field}},
  \href{https://doi.org/10.1103/PhysRevD.89.034508}{\emph{Phys. Rev. D}
  {\bfseries 89} (2014) 034508}
  [\href{https://arxiv.org/abs/1307.1509}{{\ttfamily 1307.1509}}].

\bibitem{Mahbub:2013bba}
M.S.~Mahbub, W.~Kamleh, D.B.~Leinweber and A.G.~Williams, \emph{{Searching for
  low-lying multi-particle thresholds in lattice spectroscopy}},
  \href{https://doi.org/10.1016/j.aop.2014.01.004}{\emph{Annals Phys.}
  {\bfseries 342} (2014) 270}
  [\href{https://arxiv.org/abs/1310.6803}{{\ttfamily 1310.6803}}].

\bibitem{Roberts:2013oea}
D.S.~Roberts, W.~Kamleh and D.B.~Leinweber, \emph{{Nucleon Excited State Wave
  Functions from Lattice QCD}},
  \href{https://doi.org/10.1103/PhysRevD.89.074501}{\emph{Phys. Rev. D}
  {\bfseries 89} (2014) 074501}
  [\href{https://arxiv.org/abs/1311.6626}{{\ttfamily 1311.6626}}].

\bibitem{Stokes:2013oaa}
F.M.~Stokes, W.~Kamleh and D.B.~Leinweber, \emph{{Visualizations of coherent
  center domains in local Polyakov loops}},
  \href{https://doi.org/10.1016/j.aop.2014.05.002}{\emph{Annals Phys.}
  {\bfseries 348} (2014) 341}
  [\href{https://arxiv.org/abs/1312.0991}{{\ttfamily 1312.0991}}].

\bibitem{Thomas:2014tda}
S.D.~Thomas, W.~Kamleh and D.B.~Leinweber, \emph{{Instanton contributions to
  the low-lying hadron mass spectrum}},
  \href{https://doi.org/10.1103/PhysRevD.92.094515}{\emph{Phys. Rev. D}
  {\bfseries 92} (2015) 094515}
  [\href{https://arxiv.org/abs/1410.7105}{{\ttfamily 1410.7105}}].

\bibitem{Hall:2014uca}
J.M.M.~Hall, W.~Kamleh, D.B.~Leinweber, B.J.~Menadue, B.J.~Owen, A.W.~Thomas
  et~al., \emph{{Lattice QCD Evidence that the \ensuremath{\Lambda}(1405)
  Resonance is an Antikaon-Nucleon Molecule}},
  \href{https://doi.org/10.1103/PhysRevLett.114.132002}{\emph{Phys. Rev. Lett.}
  {\bfseries 114} (2015) 132002}
  [\href{https://arxiv.org/abs/1411.3402}{{\ttfamily 1411.3402}}].

\bibitem{Owen:2015gva}
B.~Owen, W.~Kamleh, D.~Leinweber, B.~Menadue and S.~Mahbub, \emph{{Light Meson
  Form Factors at near Physical Masses}},
  \href{https://doi.org/10.1103/PhysRevD.91.074503}{\emph{Phys. Rev. D}
  {\bfseries 91} (2015) 074503}
  [\href{https://arxiv.org/abs/1501.02561}{{\ttfamily 1501.02561}}].

\bibitem{Kiratidis:2015vpa}
A.L.~Kiratidis, W.~Kamleh, D.B.~Leinweber and B.J.~Owen, \emph{{Lattice baryon
  spectroscopy with multi-particle interpolators}},
  \href{https://doi.org/10.1103/PhysRevD.91.094509}{\emph{Phys. Rev. D}
  {\bfseries 91} (2015) 094509}
  [\href{https://arxiv.org/abs/1501.07667}{{\ttfamily 1501.07667}}].

\bibitem{Trewartha:2015nna}
A.~Trewartha, W.~Kamleh and D.~Leinweber, \emph{{Evidence that centre vortices
  underpin dynamical chiral symmetry breaking in SU(3) gauge theory}},
  \href{https://doi.org/10.1016/j.physletb.2015.06.025}{\emph{Phys. Lett. B}
  {\bfseries 747} (2015) 373}
  [\href{https://arxiv.org/abs/1502.06753}{{\ttfamily 1502.06753}}].

\bibitem{Owen:2015fra}
B.J.~Owen, W.~Kamleh, D.B.~Leinweber, M.S.~Mahbub and B.J.~Menadue,
  \emph{{Transition of
  \ensuremath{\rho}\textrightarrow{}\ensuremath{\pi}\ensuremath{\gamma} in
  lattice QCD}}, \href{https://doi.org/10.1103/PhysRevD.92.034513}{\emph{Phys.
  Rev. D} {\bfseries 92} (2015) 034513}
  [\href{https://arxiv.org/abs/1505.02876}{{\ttfamily 1505.02876}}].

\bibitem{Trewartha:2015ida}
A.~Trewartha, W.~Kamleh and D.~Leinweber, \emph{{Connection between center
  vortices and instantons through gauge-field smoothing}},
  \href{https://doi.org/10.1103/PhysRevD.92.074507}{\emph{Phys. Rev. D}
  {\bfseries 92} (2015) 074507}
  [\href{https://arxiv.org/abs/1509.05518}{{\ttfamily 1509.05518}}].

\bibitem{Liu:2015ktc}
Z.-W.~Liu, W.~Kamleh, D.B.~Leinweber, F.M.~Stokes, A.W.~Thomas and J.-J.~Wu,
  \emph{{Hamiltonian effective field theory study of the $\mathbf{N^*(1535)}$
  resonance in lattice QCD}},
  \href{https://doi.org/10.1103/PhysRevLett.116.082004}{\emph{Phys. Rev. Lett.}
  {\bfseries 116} (2016) 082004}
  [\href{https://arxiv.org/abs/1512.00140}{{\ttfamily 1512.00140}}].

\bibitem{Dragos:2016rtx}
J.~Dragos, R.~Horsley, W.~Kamleh, D.B.~Leinweber, Y.~Nakamura, P.E.L.~Rakow
  et~al., \emph{{Nucleon matrix elements using the variational method in
  lattice QCD}}, \href{https://doi.org/10.1103/PhysRevD.94.074505}{\emph{Phys.
  Rev. D} {\bfseries 94} (2016) 074505}
  [\href{https://arxiv.org/abs/1606.03195}{{\ttfamily 1606.03195}}].

\bibitem{Liu:2016uzk}
Z.-W.~Liu, W.~Kamleh, D.B.~Leinweber, F.M.~Stokes, A.W.~Thomas and J.-J.~Wu,
  \emph{{Hamiltonian effective field theory study of the $\mathbf{N^*(1440)}$
  resonance in lattice QCD}},
  \href{https://doi.org/10.1103/PhysRevD.95.034034}{\emph{Phys. Rev. D}
  {\bfseries 95} (2017) 034034}
  [\href{https://arxiv.org/abs/1607.04536}{{\ttfamily 1607.04536}}].

\bibitem{Kiratidis:2016hda}
A.L.~Kiratidis, W.~Kamleh, D.B.~Leinweber, Z.-W.~Liu, F.M.~Stokes and
  A.W.~Thomas, \emph{{Search for low-lying lattice QCD eigenstates in the Roper
  regime}}, \href{https://doi.org/10.1103/PhysRevD.95.074507}{\emph{Phys. Rev.
  D} {\bfseries 95} (2017) 074507}
  [\href{https://arxiv.org/abs/1608.03051}{{\ttfamily 1608.03051}}].

\bibitem{Hall:2016kou}
J.M.M.~Hall, W.~Kamleh, D.B.~Leinweber, B.J.~Menadue, B.J.~Owen and
  A.W.~Thomas, \emph{{Light-quark contributions to the magnetic form factor of
  the Lambda(1405)}},
  \href{https://doi.org/10.1103/PhysRevD.95.054510}{\emph{Phys. Rev. D}
  {\bfseries 95} (2017) 054510}
  [\href{https://arxiv.org/abs/1612.07477}{{\ttfamily 1612.07477}}].

\bibitem{Trewartha:2017ive}
A.~Trewartha, W.~Kamleh and D.~Leinweber, \emph{{Centre vortex removal restores
  chiral symmetry}}, \href{https://doi.org/10.1088/1361-6471/aa9443}{\emph{J.
  Phys. G} {\bfseries 44} (2017) 125002}
  [\href{https://arxiv.org/abs/1708.06789}{{\ttfamily 1708.06789}}].

\bibitem{Bignell:2018acn}
R.~Bignell, J.~Hall, W.~Kamleh, D.~Leinweber and M.~Burkardt, \emph{{Neutron
  magnetic polarizability with Landau mode operators}},
  \href{https://doi.org/10.1103/PhysRevD.98.034504}{\emph{Phys. Rev. D}
  {\bfseries 98} (2018) 034504}
  [\href{https://arxiv.org/abs/1804.06574}{{\ttfamily 1804.06574}}].

\bibitem{Biddle:2018dtc}
J.C.~Biddle, W.~Kamleh and D.B.~Leinweber, \emph{{Gluon propagator on a
  center-vortex background}},
  \href{https://doi.org/10.1103/PhysRevD.98.094504}{\emph{Phys. Rev. D}
  {\bfseries 98} (2018) 094504}
  [\href{https://arxiv.org/abs/1806.04305}{{\ttfamily 1806.04305}}].

\bibitem{Wu:2018tvt}
J.J.~Wu, W.~Kamleh, D.t.~Leinweber, R.D.~Young and J.M.~Zanotti,
  \emph{{Accessing high-momentum nucleons with dilute stochastic sources}},
  \href{https://doi.org/10.1088/1361-6471/aaeb9e}{\emph{J. Phys. G} {\bfseries
  45} (2018) 125102} [\href{https://arxiv.org/abs/1807.09429}{{\ttfamily
  1807.09429}}].

\bibitem{Stokes:2018emx}
F.M.~Stokes, W.~Kamleh and D.B.~Leinweber, \emph{{Opposite-Parity
  Contaminations in Lattice Nucleon Form Factors}},
  \href{https://doi.org/10.1103/PhysRevD.99.074506}{\emph{Phys. Rev. D}
  {\bfseries 99} (2019) 074506}
  [\href{https://arxiv.org/abs/1809.11002}{{\ttfamily 1809.11002}}].

\bibitem{Stokes:2019zdd}
F.M.~Stokes, W.~Kamleh and D.B.~Leinweber, \emph{{Elastic Form Factors of
  Nucleon Excitations in Lattice QCD}},
  \href{https://doi.org/10.1103/PhysRevD.102.014507}{\emph{Phys. Rev. D}
  {\bfseries 102} (2020) 014507}
  [\href{https://arxiv.org/abs/1907.00177}{{\ttfamily 1907.00177}}].

\bibitem{Bignell:2019vpy}
R.~Bignell, W.~Kamleh and D.~Leinweber, \emph{{Pion in a uniform background
  magnetic field with clover fermions}},
  \href{https://doi.org/10.1103/PhysRevD.100.114518}{\emph{Phys. Rev. D}
  {\bfseries 100} (2019) 114518}
  [\href{https://arxiv.org/abs/1910.14244}{{\ttfamily 1910.14244}}].

\bibitem{Virgili:2019shg}
A.~Virgili, W.~Kamleh and D.~Leinweber, \emph{{Role of chiral symmetry in the
  nucleon excitation spectrum}},
  \href{https://doi.org/10.1103/PhysRevD.101.074504}{\emph{Phys. Rev. D}
  {\bfseries 101} (2020) 074504}
  [\href{https://arxiv.org/abs/1910.13782}{{\ttfamily 1910.13782}}].

\bibitem{Biddle:2019gke}
J.C.~Biddle, W.~Kamleh and D.B.~Leinweber, \emph{{Visualization of center
  vortex structure}},
  \href{https://doi.org/10.1103/PhysRevD.102.034504}{\emph{Phys. Rev. D}
  {\bfseries 102} (2020) 034504}
  [\href{https://arxiv.org/abs/1912.09531}{{\ttfamily 1912.09531}}].

\bibitem{Bignell:2020xkf}
R.~Bignell, W.~Kamleh and D.~Leinweber, \emph{{Magnetic polarizability of the
  nucleon using a Laplacian mode projection}},
  \href{https://doi.org/10.1103/PhysRevD.101.094502}{\emph{Phys. Rev. D}
  {\bfseries 101} (2020) 094502}
  [\href{https://arxiv.org/abs/2002.07915}{{\ttfamily 2002.07915}}].

\bibitem{Bignell:2020dze}
R.~Bignell, W.~Kamleh and D.~Leinweber, \emph{{Pion magnetic polarisability
  using the background field method}},
  \href{https://doi.org/10.1016/j.physletb.2020.135853}{\emph{Phys. Lett. B}
  {\bfseries 811} (2020) 135853}
  [\href{https://arxiv.org/abs/2005.10453}{{\ttfamily 2005.10453}}].

\bibitem{CSSMQCDSFUKQCD:2021rvs}
{\scshape CSSM/QCDSF/UKQCD} collaboration, \emph{{State mixing and masses of
  the \ensuremath{\pi}0, \ensuremath{\eta} and \ensuremath{\eta}' mesons from
  nf=1+1+1 lattice QCD+QED}},
  \href{https://doi.org/10.1103/PhysRevD.104.114514}{\emph{Phys. Rev. D}
  {\bfseries 104} (2021) 114514}
  [\href{https://arxiv.org/abs/2110.11533}{{\ttfamily 2110.11533}}].

\bibitem{Virgili:2022ybm}
A.~Virgili, W.~Kamleh and D.~Leinweber, \emph{{Smoothing algorithms for
  projected center-vortex gauge fields}},
  \href{https://doi.org/10.1103/PhysRevD.106.014505}{\emph{Phys. Rev. D}
  {\bfseries 106} (2022) 014505}
  [\href{https://arxiv.org/abs/2203.09764}{{\ttfamily 2203.09764}}].

\bibitem{Biddle:2022zgw}
J.C.~Biddle, W.~Kamleh and D.B.~Leinweber, \emph{{Static quark potential from
  center vortices in the presence of dynamical fermions}},
  \href{https://doi.org/10.1103/PhysRevD.106.054505}{\emph{Phys. Rev. D}
  {\bfseries 106} (2022) 054505}
  [\href{https://arxiv.org/abs/2206.00844}{{\ttfamily 2206.00844}}].

\bibitem{Biddle:2022acd}
J.C.~Biddle, W.~Kamleh and D.B.~Leinweber, \emph{{Impact of dynamical fermions
  on the center vortex gluon propagator}},
  \href{https://doi.org/10.1103/PhysRevD.106.014506}{\emph{Phys. Rev. D}
  {\bfseries 106} (2022) 014506}
  [\href{https://arxiv.org/abs/2206.02320}{{\ttfamily 2206.02320}}].

\bibitem{DBLP:journals/corr/abs-2110-09987}
C.D.~Pietrantonio, C.~Harris and M.~Cytowski, \emph{Energy-based accounting
  model for heterogeneous supercomputers}, {\emph{CoRR} {\bfseries
  abs/2110.09987} (2021) } [\href{https://arxiv.org/abs/2110.09987}{{\ttfamily
  2110.09987}}].

\bibitem{Clark:2009wm}
M.A.~Clark, R.~Babich, K.~Barros, R.C.~Brower and C.~Rebbi, \emph{{Solving
  Lattice QCD systems of equations using mixed precision solvers on GPUs}},
  \href{https://doi.org/10.1016/j.cpc.2010.05.002}{\emph{Comput. Phys. Commun.}
  {\bfseries 181} (2010) 1517}
  [\href{https://arxiv.org/abs/0911.3191}{{\ttfamily 0911.3191}}].

\bibitem{Bach:2012iw}
M.~Bach, V.~Lindenstruth, O.~Philipsen and C.~Pinke, \emph{{Lattice QCD based
  on OpenCL}}, \href{https://doi.org/10.1016/j.cpc.2013.03.020}{\emph{Comput.
  Phys. Commun.} {\bfseries 184} (2013) 2042}
  [\href{https://arxiv.org/abs/1209.5942}{{\ttfamily 1209.5942}}].

\bibitem{Boyle:2017gzg}
P.A.~Boyle, M.A.~Clark, C.~DeTar, M.~Lin, V.~Rana and A.V.~Avil\'es-Casco,
  \emph{{Performance Portability Strategies for Grid C++ Expression
  Templates}}, \href{https://doi.org/10.1051/epjconf/201817509006}{\emph{EPJ
  Web Conf.} {\bfseries 175} (2018) 09006}
  [\href{https://arxiv.org/abs/1710.09409}{{\ttfamily 1710.09409}}].

\bibitem{Bonati:2018wqj}
C.~Bonati, E.~Calore, M.~D'Elia, M.~Mesiti, F.~Negro, F.~Sanfilippo et~al.,
  \emph{{Portable multi-node LQCD Monte Carlo simulations using OpenACC}},
  \href{https://doi.org/10.1142/S0129183118500109}{\emph{Int. J. Mod. Phys. C}
  {\bfseries 29} (2018) 1850010}
  [\href{https://arxiv.org/abs/1801.01473}{{\ttfamily 1801.01473}}].

\bibitem{Cali:2021adj}
S.~Cali, W.~Detmold, G.~Korcyl, P.~Korcyl and P.~Shanahan,
  \emph{{Implementation of the conjugate gradient algorithm for heterogeneous
  systems}}, \href{https://doi.org/10.22323/1.396.0507}{\emph{PoS} {\bfseries
  LATTICE2021} (2022) 507} [\href{https://arxiv.org/abs/2111.14958}{{\ttfamily
  2111.14958}}].

\bibitem{Yamaguchi:2022feu}
A.~Yamaguchi, P.~Boyle, G.~Cossu, G.~Filaci, C.~Lehner and A.~Portelli,
  \emph{{Grid: OneCode and FourAPIs}},
  \href{https://doi.org/10.22323/1.396.0035}{\emph{PoS} {\bfseries LATTICE2021}
  (2022) 035} [\href{https://arxiv.org/abs/2203.06777}{{\ttfamily
  2203.06777}}].

\end{thebibliography}

\providecommand{\href}[2]{#2}\begingroup\raggedright\endgroup

\end{document}